\documentclass{aastex}

\shortauthors{Sil'chenko}
\shorttitle{NGC 524 and NGC 6340}

\begin{document}

\title{Face-on galaxies NGC 524 and NGC 6340: chemically
 decoupled nuclei and inclined circumnuclear disks}

\author{O. K. Sil'chenko\altaffilmark{1}}
\affil{Sternberg Astronomical Institute, Moscow, 119899 Russia\\
       Isaac Newton Institute, Chile, Moscow Branch\\
     Electronic mail: olga@sai.msu.su}

\altaffiltext{1}{Guest Investigator of the RGO Astronomy Data Centre}

\begin{abstract}

The central regions of the early-type disk galaxies NGC~524 and NGC~6340
have been investigated with the Multi-Pupil Field Spectrograph
at the 6m telescope of the Special Astrophysical Observatory of
the Russian Academy of Sciences. We confirm the existence of
chemically distinct stellar nuclei in these galaxies which have been
claimed earlier. By myself, the metallicity differences which are found
between the nuclei and the bulges, 0.5--1.0~kpc from the centers, reach
0.5--0.6 dex. Both nuclei are magnesium overabundant, but the bulges
have different magnesium-to-iron ratios: it is solar in NGC~6340 and the
same as the nuclear one in NGC~524. The kinematical and morphological
analyses reveal the existence of inclined central disks
in these galaxies. In NGC~524 the central disk consists of stars, dust,
and ionized gas; its extension may be as large as up to $R\approx 3$ kpc,
and it is inclined by some 20\degr\ to the global galactic plane.
In NGC~6340 only a gaseous polar disk is detected which extension
does not exceed $R\approx 500$ pc.

\end{abstract}


\section{Introduction}

Early-type disk galaxies differ from late-type disk (spiral)
galaxies by a smooth appearance of their disks: in general, they
lack a massive gaseous component and noticeable star formation
regions. However, often some gas is present in the circumnuclear
regions; it is widely accepted that this gas has an external
origin. Bertola with co-workers have reported a large statistics
of decoupled distributions and kinematics of gaseous and stellar
components in nearby lenticulars. Particularly, \citet{b92a}
have reported three cases of gas counterrotation in a sample
of 15 objects and concluded, by involving literature data,
that at least 40\%\ gas-possessing S0 have to accrete this gas
during their late evolution. \citet{kfm96} also noted that
a significant fraction of gas disks in lenticulars, $24\% \pm 8\%$
in their sample, demonstrate counterrotation with respect to stars.
Among the known decoupled distributions of gas and stars,
a class of polar rings is the most spectacular one. Besides the extreme
case of NGC~2685 where the radius of the gaseous polar
ring is comparable to the radius of the global stellar disk, there
exist more "minor-axis dust-lane" galaxies, such as
NGC~1947 \citep{b92b} or NGC~7280
\citep{cszm97,we7280} where compact circumnuclear gaseous disks
are shown to rotate orthogonally with respect to the global stellar
rotation.

In 1992 we have published our first paper \citep{we92} on chemically
distinct nuclei in early-type disk galaxies. In 3 lenticulars and
in 3 Sa--Sb's we have found a sharp drop of magnesium absorption line
strength when passing from the nuclei to the surrounding bulges.
The difference in Mg absorption strength indicates an order of
magnitude change in metallicity on scales of a few arcseconds whereas
smooth metallicity gradients in galactic spheroids are usually only
0.3 dex per a decade in radius in ellipticals \citep{cdb93} or somewhat
larger, with a mean of --0.5 and an extremum of --1.0 per dex, in
bulges of early-type disk galaxies \citep{bp94,fish96}.
The origin of chemically decoupled nuclei in disk galaxies may be
related to a gas accretion event and a subsequent star formation burst
in the nucleus where the accreted gas is accumulated. With
this hypothesis we try to find a connection between the chemical
distinctness of the nuclear stellar population and the presence
of gas subsystems of obviously external origin. We \citep{we92,sil99a}
have already noted chemically distinct nuclei in the lenticulars
NGC~1023 where an inclined extended HI disk is present
and NGC~7332 where circumnuclear gas counterrotates
with respect to the stars. The lenticular galaxy with the chemically
decoupled nucleus NGC~7280 represents a remarkable example:
the circumnuclear gas coupling with tiny dust lanes is extended
and rotates orthogonally to the circumnuclear stellar disk which
is, on its own, inclined with respect to the main galactic disk
\citep{we7280}.

In this paper we present the results of a complex study for the
central parts of two nearby face-on early-type disk galaxies:
NGC~524 and NGC~6340. Some time ago we
found chemically decoupled nuclei in these galaxies.
NGC~524 \citep{we92} and NGC~6340
\citep{sil95} were observed at the 6m telescope
with the Multi-Pupil Field Spectrograph (MPFS) equipped by IPCS, and
the drops of the magnesium absorption line equivalent width along
the radius were detected. Recently we have re-observed the
galaxies because the MPFS is now equipped with CCD and the data
becomes more accurate. Besides the investigation of the radial
variations of absorption-line indices, we present also results
of kinematical and morphological analyses for the central parts
of NGC~524 and NGC~6340. Due to face-on
orientation of their main planes, any inclined disks, both
stellar and gaseous, can be easily detected if systematic visible
velocity gradients are observed which can be attributed to rotation
velocity projection onto the line of sight. The paper is organized
as follows. Section~2 describes the observations and their reduction
as well as the data from open archives which we use. In Section~3
we re-consider radial variations of metal absorption-line indices
and confirm that the stellar nuclei of NGC~524 and
NGC~6340 are chemically decoupled. In Section~4 we study
the morphology of distributions of gas and stars in the galaxies,
and in Section~5 we present two-dimensional velocity fields both
for ionized gas and stars in the central regions. Section~6 contains
the main conclusions. In Table~1 we give the basic parameters for the
galaxies under consideration.

\begin{table}
\caption[ ] {Global parameters of the galaxies}
\begin{flushleft}
\begin{tabular}{lcc}
\hline\noalign{\smallskip}
NGC & 524 & 6340 \\
\hline\noalign{\smallskip}
Type (NED) & SA(rs)$0^+$ & SA(s)0/a  \\
$R_{25}$, kpc (RC3+LEDA) & 13.2 & 9.3 \\
$B_T^0$ (RC3) & 11.17 & 11.67  \\
$M_B$ (LEDA) & --21.39 & --19.94 \\
$(B-V)_T^0$ (RC3) & 1.00 & 0.79 \\
$(U-B)_T^0$ (RC3) & 0.58 &  \\
 $V_r $ (RC3) & (opt) 2421 $km\cdot s^{-1}$ & (radio)
      1198 $km\cdot s^{-1}$  \\
Distance, Mpc (LEDA, $H_0$=75 $km\cdot s^{-1}\cdot Mpc^{-1}$) & 33
    & 19.8 \\
Inclination (LEDA) & $8.7^o$ & $25.6^o$  \\
$PA_0$ (LEDA) &  & $120^\circ$  \\
$\sigma_0$, km/s (LEDA) & 246 & 146\\
$v_m$, km/s (LEDA) & & 245\\
\hline
\end{tabular}
\end{flushleft}
\end{table}

\section{Observations and Data Reduction}

The observations of NGC~524 and NGC~6340 which we present in this paper
have been performed in 1996--1997 with the Multi-Pupil Field Spectrograph
(MPFS) of the 6m telescope in the Special Astrophysical Observatory,
Nizhnij Arkhyz, Russia \citep{afetal90,afman}. The detailed log of
the observations is given in Table~2. The detector used was
$520 \times 580 $ CCD ISD015A ("Electron", St-Petersburg), with
the pixel size $18\mu m \times 24 \mu m$ and read-out noise of 13$e^-$.
The gain was 3.4$e^-$ in 1996 and 7$e^-$ in 1997.

\begin{table}
\caption[ ] {2D spectroscopy of NGC~524 and NGC~6340}
\begin{flushleft}
\begin{tabular}{lllccc}
\hline\noalign{\smallskip}
Date & Galaxy & Exposure &
Sp. range, \AA\ & PA(top) & Seeing \\
\hline\noalign{\smallskip}
14/15.08.96 & NGC~6340 & 90 min &
4800--5400 & $209\arcdeg$ & $1\farcs 7 $ \\
14/15.08.96 & NGC~6340 & 90 min &
 6250--6900 & $180\arcdeg$ & $1\farcs 7 $ \\
14/15.10.96 & NGC~524 & 60 min &
 6250--6900 & $67\arcdeg$ & $1\farcs 6 $ \\
31.10/1.11.97 & NGC~6340 & 81 min &
 4800--5400 & $164\arcdeg$ & $2\farcs 2 $ \\
31.10/1.11.97 & NGC~524 & 94 min &
 4800--5400 & $149\arcdeg$ & $2\farcs 2 $ \\
\hline
\end{tabular}
\end{flushleft}
\end{table}

MPFS which resembles the French integral-field spectrograph TIGER
\citep{betal95} by the principles of its design allows to obtain
simultaneously a set of spectra from an extended area ($8 \times 12$
square elements in the present work); each spatial element is
$1\farcs 3 \times 1\farcs 3$, so the full field of view which we
centers onto the nuclei of the galaxies is $10\arcsec \times 16\arcsec$.
Two spectral ranges were exposed, blue and red, with a (reciprocal)
dispersion of 1.6~\AA\ per pixel (the spectral resolution of
4--6~\AA, slightly varying over the field of view). The blue
spectral range under consideration contains several strong absorption
lines such as the Mgb feature. We have used it to derive stellar
velocity fields by cross-correlating spectra of the galaxies with the
spectra of K giant stars observed during the same nights with the same
spectrograph. Also we have calculated absorption-line indices
H$\beta$, Mgb, and Fe5270 in the popular Lick system \citep{woretal}
to study radial variations of the mean metallicity of the stellar
populations. The sky background in the blue was exposed separately
after each galaxy exposure. Properly normalized and smoothed, it was
then subtracted from the spectra of the galaxies. The exposure times
for the galaxies were long enough to achieve a signal-to-noise ratio
not less than 50--70 per \AA\ in the nuclei -- this level of S/N
provides an accuracy
of the absorption-line indices better than 0.2~\AA\ \citep{cgcg}.
However the accuracy drops to 0.5--0.6~\AA\ for the outermost
elements observed, and to keep a constant level of signal-to-noise
ratio up to $R\approx 8\arcsec$ we added the element spectra
in concentric circular rings with the width of $1\farcs 3$ and
the center in the nuclei. After azimuthal averaging, the accuracy
of all indices measured along the radius is about 0.1~\AA. To check
consistency of our index system with the standard Lick one, we have
observed 9 G8--K3 stars, both giants and dwarfs, from the list of
\citet{woretal} and have calculated their indices
in the same manner as those for the galaxies. The agreement between
stellar indices measured by us and those tabulated in \citet{woretal}
is excellent within the errors cited by them; the mean deviations of
our measurements from those of \citet{woretal} for
all the indices are less than 0.05~\AA, so we need no degenerate
our spectral resolution to come into agreement with the standard
Lick index system. The indices Mgb and Fe5270 measured in the galaxies
have been corrected for the stellar velocity dispersion broadening; the
correction values are determined by artificial Gauss broadening the
stellar spectra with varying $\sigma$'s. For NGC~6340 this
correction is small, only 0.1~\AA, because the stellar velocity
dispersion in the center of this galaxy is 100--140~km/s
\citep{bottema}. For NGC~524 this correction is
0.5~\AA\ in the nucleus and 0.4~\AA\ beyond it, because the nuclear
stellar velocity dispersion is 246~km/s (LEDA) or $236 \pm 19$~km/s
\citep{schechter}, and it decreases only slightly in the nearest
vicinity of the nucleus (our impression from the visual inspection of
the spectra). The red spectral range under consideration contains
the emission line [NII]$\lambda$6583, strong in the centers of
NGC~524 and NGC~6340. It is used to obtain two-dimensional velocity
fields of the ionized gas. The wavelength calibration for both spectral
ranges was made by using separate exposures of the hollow-cathod
lamp filled with helium, neon, and argon. The accuracy and absence
of systematic shifts of the velocity scale was checked by measuring
the night-sky emission lines. The accuracy of the individual velocity
measurements both for stars and ionized gas is about 25 km/s.
The primary data reduction -- bias subtraction, flatfielding,
cosmic ray removal, extraction of one-dimensional spectra,
wavelength calibration, construction of surface brightness maps --
have been performed with the software developed in the Special
Astrophysical Observatory \citep{vlas}. The absorption-line
indices were calculated with our own FORTRAN programs.

\begin{table}
\caption[ ] {Photometric observations of NGC~524 and 6340}
\begin{flushleft}
\begin{tabular}{ccccccc}
\hline\noalign{\smallskip}
Date & Galaxy & Telescope & Filter & Exposure & Seeing & Scale \\
\hline\noalign{\smallskip}
9/10.05.88 & NGC~6340 & JKT & $V$ & 600 s
& $1\farcs 6$ & $0\farcs 30$\\
9/10.05.88 & NGC~6340 & JKT & $R$ & 600 s
& $1\farcs 7$ & $0\farcs 30$\\
9/10.05.88 & NGC~6340 & JKT & $I$ & 720 s
& $1\farcs 6$ & $0\farcs 30$\\
17/18.05.98 & NGC~6340 & JKT & $R$ & 900 s
& $1\farcs 1$ & $0\farcs 33$\\
6/7.11.96 & NGC~524 & JKT & $I$ & 360 s
& $2\farcs 6$ & $0\farcs 31$ \\
24.09.95 & NGC~524 & HST+WFPC2 & F555W & 160 s & $0\farcs13$ &
$0\farcs 045$ \\
24.09.95 & NGC~524 & HST+WFPC2 & F814W & 320 s & $0\farcs13$ &
$0\farcs 045$ \\
2.06.96 & NGC~6340 & HST+WFPC2 & F606W & 400 s & $0\farcs15$ &
$0\farcs 045$ \\
\hline
\end{tabular}
\end{flushleft}
\end{table}

We have also retrieved a long-slit spectrum of NGC~524 in the La Palma
Archive. The galaxy was observed on December 24, 1995, at the
ISIS WHT with the dispersion of 0.4~\AA\ per pixel in the spectral
range 5000--5400~\AA\ (blue arm of the spectrograph); the slit
(width equal to $0\farcs 62$) was aligned almost along the minor
axis of the galaxy,  $PA=135\arcdeg$. We have used this spectrum
to calculate Mgb and Fe5270 indices along the slit with the same
techniques as the indices from our MPFS data (the sky for subtracting
was taken from the edges of the slit) though we
cannot reduce these indices into the standard Lick system because
of the lack of standard Lick star observations.

The photometric data involved in our analysis are
taken from the La Palma and HST Archives. The details of
the observations are given in Table~3.
The programs in the frame of which the central parts of the galaxies
have been observed at the Hubble Space Telescope are
"Core properties of bulges of spiral galaxies"
(Principal Investigator: Stiavelli, Program ID:
6359) and "Nuclear structure of S0 galaxies"
(Principal Investigator: Phillips, Program ID: 5999).
We have derived morphological characteristics of the surface
brightness distribution in the galaxies by analysing these images;
the program FITELL of Dr.~Vlasyuk has been used for this purpose.

\section{Chemically decoupled nuclei in NGC~524 and 6340}

As we have noted in the Introduction, the galaxies under consideration
exhibit stellar nuclei with enhanced magnesium absorption
lines: for NGC~524, see \citet{we92},
for NGC~6340, see \citet{sil95}. Now we present more
precise radial profiles of Mgb together with those of Fe5270 and
H$\beta$. The azimuthally averaged index measurements are presented
in Table 4. The accuracy of the azimuthally averaged measurements is
improved by a factor of 3 with respect to our earlier results.
However, the qualitative conclusions remain the same.

\begin{table}
\caption{Azimuthally averaged Lick indices in NGC~524 and NGC~6340}
\begin{tabular}{l|ccc|ccc|ccc}
\hline
$R\arcsec$ &
\multicolumn{3}{|c|}{NGC 524} &
\multicolumn{3}{|c|}{NGC 6340, Aug96} &
\multicolumn{3}{|c}{NGC 6340, Oct97}\\
\cline{2-10}
 &
H$\beta$ & Mgb & Fe5270 &
H$\beta$ & Mgb & Fe5270 &
H$\beta$ & Mgb & Fe5270 \\
\hline
0   &  1.33 & 4.87 & 3.06  &  1.05 & 4.65 & 2.68 & 1.56 & 4.49 & 2.93\\
1.3 &  1.42 & 4.65 & 2.97  &  1.24 & 4.20 & 2.88 & 1.35 & 4.38 & 2.83\\
2.6 &  1.23 & 4.20 & 2.75  &  1.30 & 3.60 & 2.43 & 1.00 & 3.83 & 2.48\\
3.9 &  1.16 & 3.88 & 2.34  &  0.91 & 2.97 & 2.34 & 1.23 & 3.18 & 2.34\\
5.2 &  0.94 & 3.70 & 2.27  &  0.80 & 3.16 & 2.14 & 1.24 & 3.17 & 2.30\\
6.5 &  1.22 & 3.68 & 2.21  &  0.92 & 2.26 & 1.64 & 1.39 & 3.10 & 2.00\\
7.8 &  0.98 & 3.61 &  & 0.78 & 2.24 & 2.03 & 1.59 & 2.49 & 0.99\\
\hline
\end{tabular}
\end{table}

NGC~524 (Fig.~1) shows radial profiles of the magnesium and iron
indices of very similar shapes. The maximum equivalent widths
of the metal lines are observed in the nucleus and decrease radially
up to $R\approx 4\arcsec$ (600 pc). Between $R=4\arcsec$ and
$R=8\arcsec$ the MPFS profiles flatten off. Their extensions by the
long-slit data confirm the flat behavior of the metal-index profiles
in the radius range $5\arcsec - 20\arcsec$ (the bulge-dominated region)
though the long-slit measurements are not reduced into the Lick system.
By calculating index gradients in the bulge in a traditional manner,
versus $\log R$ in the radial range $3\arcsec - 8\arcsec$,
we have obtained
$d\mbox{Mgb} / \log R =-0.85 \pm 0.18$\AA/dex and
$d\mbox{Fe5270} / \log R =-0.59 \pm 0.02$\AA/dex,
both gradients corresponding to $d\mbox{[m/H]} / \log R =-0.4$ --
a value quite typical for early-type galaxies \citep{bp94,fish96}.
The zero-points of the linear regressions corresponding to
the extrapolated bulge indices at $R=1\arcsec$, Mgb=$4.34 \pm 0.14$~\AA\
and Fe5270=$2.68 \pm 0.01$~\AA, are noticeable smaller than
the nuclear indices, $\mbox{Mgb(nuc)}=4.87 \pm 0.16$~\AA\ and
$\mbox{Fe5270(nuc)}=3.16 \pm 0.19$~\AA.
We have taken the mean bulge characteristics by averaging the MPFS
data in the radial range $4\arcsec - 8\arcsec$, resulting in
$<\mbox{Mgb(bul)} >=3.75 \pm 0.06$~\AA\ and
$<\mbox{Fe5270(bul)} >=2.27 \pm 0.04$~\AA.
If we apply now the models for old
stellar populations of \citet{worth94} to the differences $\Delta$Mgb
and $\Delta \mbox{Fe5270}$ between the nucleus and the mean bulge
under the assumption of equal mean stellar population ages, we
obtain $\Delta$[Fe/H]=$0.50 \pm 0.10$ dex
from $\Delta$Mgb and $\Delta$[Fe/H]=$0.56 \pm 0.15$ dex from
$\Delta \mbox{Fe5270}$, the nucleus being more metal-rich.
Unfortunately, we cannot determine the mean age of the stellar
population in the center of NGC~524 by using the
absorption-line index H$\beta$ because it is notably affected
by the emission line. However, usually chemically decoupled
nuclei in lenticular galaxies appear to be younger than the bulges
\citep{sil99a,we7280}, so the derived metallicity difference
must be considered as a low limit. The coincidence, within the errors,
of the $\Delta$[Fe/H] values derived from  $\Delta$Mgb and
$\Delta \mbox{Fe5270}$ gives some evidence for equal magnesium-to-iron
ratios in the nucleus and in the bulge of NGC~524.

NGC~6340 (Fig.~2) has been twice observed in the green spectral range
with the MPFS+CCD, so we can independently estimate the accuracy of
our measurements from the comparison of two data sets. One can see in
Fig.~2 that the claimed accuracy of 0.1~\AA\ for the azimuthally
averaged points is kept quite well up to $R\approx 6\arcsec$. The shape
of the radial profiles of the metal-line indices demonstrates again
the presence of a chemically decoupled nucleus though the contrast
of the Fe5270 drop is obviously smaller than that of Mgb.
As we have in this galaxy more measurements than in NGC~524,
the mean indices for the nucleus and the bulge of NGC~6340
are estimated more precisely: for the nucleus,
$\mbox{Mgb(nuc)}=4.57 \pm 0.08$~\AA\ and
$\mbox{Fe5270(nuc)}=2.80 \pm 0.13$~\AA, and for the bulge,
$R=4\arcsec-6\arcsec$, $\mbox{Mgb(bul)}=3.22 \pm 0.04$~\AA\ and
$\mbox{Fe5270(bul)}=2.32 \pm 0.07$~\AA. After applying the models
of \citet{worth94} to the nucleus-bulge differences $\Delta$Mgb and
$\Delta \mbox{Fe5270}$, we obtain $\Delta$[Fe/H]=$0.61 \pm 0.05$ dex
from $\Delta$Mgb and $\Delta$[Fe/H]=$0.30 \pm 0.13$ dex from
$\Delta \mbox{Fe5270}$. Again, these metallicity differences
represent only low limits due to age uncertainty: a narrow H$\beta$
emission line is seen inside the broader absorption and so the
absorption-line index H$\beta$ is invalid for the age diagnostics.
Unlike NGC~524, NGC~6340 demonstrates
different $\Delta$[Fe/H] from the magnesium and iron index differences;
it may be a signature of different magnesium-to-iron ratios in the
bulge and in the nucleus of this galaxy.

The question on magnesium-to-iron ratio can be addressed directly
with a diagram (Fe5270, Mgb). The first fruitful attempt to use
this diagram to diagnose the Mg/Fe ratio was made by \citet{wor92},
and they found immediately that a bulk of luminous ellipticals
are magnesium overabundant. I noted that Mgb and Fe5270 in the
centers of disk galaxies, including lenticulars, satisfy the
models with solar magnesium-to-iron ratio \citep{me93b}, though
later \citet{fish96} reported several nuclei (but not the bulges!)
in luminous lenticulars which were magnesium overabundant. Figure~3
presents these diagrams for NGC~524 and NGC~6340; the model sequences
with varying metallicity and ages calculated by \citet{worth94} for
[Mg/Fe]=0 border the narrow locus of the solar Mg/Fe ratio. The
comparison of the observations with the models reveals that both
chemically decoupled nuclei are magnesium overabundant; though less
accurate, the measurements of \citet{trager} confirm this conclusion.
The magnesium overabundance of the chemically decoupled nuclei in
lenticular galaxies is not unique: we have found it for example in
NGC~1023 \citep{sil99a}; but it is not a rule --
NGC~7332 and NGC~7280, being of the same
total luminosity as NGC~6340, demonstrate the solar Mg/Fe in
their chemically decoupled nuclei as well as in their bulges
\citep{sil99a,we7280}. NGC~6340 shows solar magnesium-to-iron
ratio in its bulge; the [Mg/Fe] difference between the nucleus and
the bulge is about 0.3 dex.
NGC~524 demonstrates a quite different Mg/Fe behavior: the data
taken along the radius lie parallel to the model sequence but shifted
to the right. This is interpreted as $\mbox{[Mg/Fe]}\approx +0.3$
constant up to $R\approx 8\arcsec$ (1 kpc) in this galaxy. Such behavior
of Mg/Fe resembles that in ellipticals \citep{wor92} and in some bulges
of supergiant spirals like M~31 \citep{silbv98}
or NGC~488 \citep{sil99b}.

\section{Morphology of stellar and gas distributions in the center
of NGC~524 and NGC~6340}

To understand the structure of the central regions in NGC~524 and
NGC~6340, we have analysed digital images obtained
from the La Palma and HST Archives. The radial dependencies of the
major axis position angle and ellipticity of the isophotes approximated
by ellipses are presented in Fig.~4 for NGC~524 and
in Fig.~5 for NGC~6340.

The galaxies have been earlier studied photometrically, especially
NGC~524. For NGC~6340 only one-dimensional
brightness profiles can be found in the literature \citep{bor,wk82};
they give evidence that the global disk begins to dominate at
$R \geq 15\arcsec$. For NGC~524 the information is rich
and somewhat controversial. According to \citet{kent85} and \citet{bg90},
which have made the brightness profile decomposition in the $r$-band
and $B$-band respectively, the bulge dominates over the disk at all
radii; so in some respect NGC~524 may be re-classified
as an elliptical with embedded disk. However, the profiles presented
by \citet{hs76} and \citet{ital92} reveal an extended exponential disk
dominating over the bulge at $R \geq 30\arcsec$. The ellipticity
estimates for the outer isophotes range from $0.03 \pm 0.01$
\citep{hs76} to 0.06 \citep{kent84}; our estimate from the La Palma data
is 0.06; in any case, the galaxy inclination
cannot be larger than $20\degr$. The line-of-nodes orientation lies
somewhere between $PA_0=5\degr \pm 23\degr$ \citep{hs76} and $41\degr$
\citep{kent84,ital92} -- more exact determination is prevented by
the low ellipticity of the isophotes, particularly, LEDA and RC3 omit
the estimate of the major-axis position angle (see our Table~1).

The combination of the photometric data of different spatial resolution
allows us to study the structure of the inner regions of
NGC~524 and NGC~6340 in more detail.

Figure~4 shows the radial variations of the major-axis position angle
and ellipticity in the center of NGC~524. At $R > 5\arcsec$
they look rather constant, $PA_0$ at the level of about $40\degr$ and
$1-b/a$ at the level of 0.03. This behavior does not contradict
the data for the outer regions according e.g. to \citet{ital92}.
However, at $R \leq 4\arcsec$ the major axis seems to be turned
by some $20\degr$ and, more important, the ellipticity seems to increase
up to about 0.10 -- the value which is not reached even in the outermost
part of the galaxy. Though an accuracy of the morphological parameter
estimates is low for such roundish isophotes and though some dust is
present inside $R \approx 1\arcsec$ \citep{hst2} which causes
a discrepancy of two HST measurements of the ellipticity through the
different filters , a presence of the
visibly elongated stellar structure in the center of
NGC~524 seems to be quite probable.

The variations of the morphological parameters in the center of
NGC~6340 are even more prominent than in the center of
NGC~524. In the disk-dominated region, at
$R=30\arcsec -35\arcsec$, the position angle of the isophote major
axis, that is, the orientation of the line of nodes, is $PA_0=131\degr$.
The asymptotic ellipticity is 0.08, so the inclination of the
galactic plane may be as large as $23\degr$. However, at $R < 5\arcsec$
(Fig.~5) the major axis is surely turned and its $PA\approx 85\degr$
differs from the orientation of the line of nodes at least by $45\degr$.
The ellipticity demonstrates a local maximum at $R\approx 5\arcsec$,
the most prominent in the high-resolution HST data.
Though some discrepancy of the data inside $R\approx 3\arcsec$ which
is caused by the different spatial resolutions of the observations
can be seen, the elongated stellar structure in the center of
NGC~6340 is also rather probable.

Our spectral observations have detected the emission line
[NII]$\lambda$6583, and so the presence of ionized gas, in the
centers of both galaxies. We wonder if the distributions of diffuse
matter, gas and dust, resemble those of the stellar components. First,
we can examine the dust. Figure~6 presents direct WFPC2 (namely PC)
images of both galaxies. One can see a rather extended roundish
dust disk in the center of NGC~524 which consists of tightly wrapped
thin dark spirals. Lower-resolution observations made by \citet{vcv88}
also revealed a red nucleus and "almost face-on dust ring" with
the sizes of $16\arcsec \times 12\arcsec$; let us note, however,
that such axis ratio, 0.75, favors a dust disk inclination rather
close to $40\degr$, not exactly face-on. In the center of
NGC~6340 (Fig.~6b) \citet{cszm97} noticed "a tiny dust
lane". We can add that the very compact, with the radius of less
than $0\farcs 5$, dust disk is seen almost edge-on
($i_{dust} \geq 70\degr$), and therefore we deal with a circumnuclear
polar ring. Its line of nodes is close to $PA_0\approx 90\degr$.

Our MPFS observations allow to refer directly to emission-line
brightness maps. Figure~7 presents surface distributions of the
[NII] emission intensity in arbitrary units. In NGC~6340
the nitrogen emission-line isophotes are elongated in
$PA\approx 130\degr$, so they imply the existence of rather extended
gaseous disk which line of nodes, but not the inclination, coincides
with that of the global disk. In NGC~524 the emission
distribution is strongly asymmetric: it is confined to the western
part of the circumnuclear region, and its geometry is not quite
evident. However, if we remind the work of \citet{macetal} who observed
lenticular galaxies through the narrow-band
$\mbox{H}_\alpha + \mbox{[NII]}$ filter, we find that the emission
distribution in the center of NGC~524 matches perfectly
that of the dust: the same roundish patchy disk with the radius of
some $20\arcsec$. A bright emission knot in the western vicinity of
the nucleus is also seen in the map presented by \citet{macetal}, so
our Fig.~7a represents probably a high-level slice of the overall
emission distribution.

\section{Kinematics of gas and stars in the central regions
of NGC~524 and 6340}

When we look at a rotating disk from its pole, it is projected
"face-on" (on the sky plane) and the projection of its rotation velocity
onto the line of sight is zero. Generally, face-on galaxies should
lack any line-of-sight velocity gradients (over their images).
It is just what we would expect from NGC~524 and
NGC~6340. However the line-of-sight velocity fields
obtained by us for these galaxies look quite different.

Figure~8 presents velocity fields for the ionized gas in the centers
of the galaxies. Both maps demonstrate obvious signs of regular
rotation: the measured line-of-sight velocities gradually changes
from one map corner to another. In the case of axisymmetrical
(circular, cylindric) rotation the direction of the highest
velocity gradient (which we call "dynamical major axis") should
coincide with the line of nodes. In NGC~524 (Fig.~8a)
this direction is $PA_0\approx 23\degr$, in NGC~6340
(Fig.~8b) -- $PA_0\approx 115\degr$. Both directions are indeed
close to the major axes of the inner isophotes as measured by us
in the previous Section, though it is not quite true for
NGC~6340: $PA_0(phot)$ changes along the radius
in the center of this galaxy, and it is not quite clear what orientation
must be chosen as a reference one. But as the first approximation,
one can conclude that we see circumnuclear rotating
thin (cold) gaseous disks; and these disks, unlike the main stellar
disks, are obviously not face-on: their visible rotation is very
fast.

As for the stellar rotation, we have not found clear signs of it
in the center of NGC~6340. The systematic velocity
variation over the full line-of-sight velocity field of this
galaxy, the spatial base of which is some $10\arcsec$, if it exists,
does not exceed the random error of one velocity measurement, namely,
25~km/s. On the contrary, stars in the center of NGC~524
demonstrate quite noticeable rotation (Fig.~9). This is somewhat
unexpected: LEDA (see Table~1) gives for NGC~524
$i=8.7\degr$ and for NGC~6340 $i=25.6\degr$, so the
latter galaxy would have a projection factor three times less than
the former. Besides, the stellar velocity dispersion in the
center of NGC~524 is rather high, the galaxy is thought
by many photometrists to be bulge-dominated, so this galaxy is closer
to ellipticals than NGC~6340, and as a luminous early-type
galaxy it should rotate slowly. However, the line-of-sight velocity
gradient is quite visible in Fig.~9; the direction of the maximum
velocity gradient, $PA_0\approx 19\degr$, agrees well with the
dynamical major axis of the gaseous component. As the photometric
major axis in the radius range of $1\arcsec - 4\arcsec$ also is
aligned in $PA_0(phot)\approx 20\degr$ (Fig.~4), we conclude that
the separate stellar disk with the radius of $4\arcsec - 5\arcsec$
(0.6--0.8 kpc) having a gaseous extension up to $R\approx 20\arcsec$
(3 kpc) exists in the center of NGC~524. Its line
of nodes is aligned in $PA\approx 20\degr$, whereas the global
line of nodes is close to $PA_0\approx 40\degr$ (Fig.~4, see also
\citet{ital92}), so the circumnuclear disk is inclined with
respect to the main galactic plane.

To compare more thoroughly stellar and gaseous rotations in the
centers of NGC~524 and NGC~6340, we have
simulated one-dimensional velocity profiles along the dynamical
major axes overlaping a narrow "slit" on the two-dimensional velocity
fields for the gas and stars. The resulting profiles are plotted
in Fig.~10. In NGC~524 (Fig.~10a) there is no significant
discrepance between the stellar and gaseous rotation; however both
velocity variation amplitudes are too high to be attributed to the
rotation plane inclined by $i \leq 20\degr$, unless we accept
$V_{rot} \geq 360$~km/s at $R < 1$ kpc. We would rather ascribe
$i\approx 40\degr$ implied by the morphology of the dust ring
\citep{vcv88} to this fast-rotating circumnuclear disk. This is
one more sign for its decoupling. In NGC~6340
we had a problem with determining a position angle for our
cross-section: the ionized gas shows $PA_0 \approx 115\degr$,
the dynamical major axis of the circumnuclear stellar component
is unknown, and the global line of nodes is close to
$PA_0 \approx 140\degr$ (Fig.~5). \citet{bottema} has chosen
$PA=130\degr$ for his long-slit observations, and to make also
a comparison with his data we cut the velocity fields along
$PA=130\degr$. The gas and stellar velocity profiles in the center
of NGC~6340 look quite different (Fig.~10b): the
slope of the gas velocity profile, though it is not taken exactly
along the dynamical major axis, is at least five times steeper
than that for the stellar component. The central stellar velocity
dispersion is low; galactic bulges with such $\sigma_*$ usually
rotate rapidly \citep{ki82}.
An explanation which can reconcile the visible kinematics
of both components is that the gas and stars in the center
of NGC~6340 rotate in different planes. While
\citet{bottema} decided that the visible stellar rotation of
NGC~6340 is quite normal for the inclination of
$20\degr$, the rotation of the ionized gas must be more related to
the edge-on circumnuclear dust ring
visible on the image provided by the HST (Fig.~6b).
The highly elongated isophotes of the [NII] emission brightness
surface distribution (Fig.~7b) support this suggestion.

\section{Conclusions and Discussion}

Two early-type face-on disk galaxies, NGC~524 and
NGC~6340, where we previously suspected the presence
of chemically decoupled nuclei, have been re-investigated with
the Multi-Pupil Field Spectrograph of the 6m telescope equipped
with CCDs. We confirm a drop of the magnesium-line index between
the nuclei and the bulges $4\arcsec - 8\arcsec$ from the centers
in these galaxies; the $\Delta$Mgb's measured now imply metallicity
differences of 0.5--0.6 dex. In NGC~524 the iron-line
index Fe5270 also demonstrates a drop between the nucleus and the
bulge; the magnesium overabundance, [Mg/Fe]$\approx +0.3$, is
almost constant over the full radius range under consideration.
In NGC~6340 the nucleus is magnesium overabundant too,
but the ratio [Mg/Fe] falls to zero toward $R\approx 4\arcsec$.

The presence of chemically decoupled nuclei in the galaxies is
accompanied by indications of inclined disks detected in the
circumnuclear regions though our investigation is completely confined
to the bulge-dominated regions. In NGC~524 the central velocity
fields of the ionized gas and stars are similar; the orientation
of the dynamical major axis, $PA_0 \approx 20\degr$, coincides
with the photometric major axis orientation in the radius range of
$1\arcsec - 5\arcsec$. Though both deviate from the line of nodes
which has $PA_0=32\degr - 42\degr$, the coincidence of the
photometric and dynamical major axes proves the axisymmetric
character of rotation: in a triaxial potential
the photometric and dynamical major axes turn in opposite directions
with respect to the line of nodes \citep{mbe92}. As the gaseous
and stellar rotation velocities are comparable, probably we deal
with a stellar-and-gaseous disk inclined with respect to the main
plane of the galaxy. The rather high visible velocity variation
amplitude favors larger inclination of the circumnuclear disk than
is that of the global disk of the galaxy: from the geometry of
the dust distribution \citep{vcv88} we would guess about
$i\approx 40\degr$ in the center versus $i\leq 20\degr$ for the
whole galaxy. The central disk may be extended up to
$R\approx 20\arcsec$ (3 kpc) as the dust spirals and emission-line
distribution in \citet{macetal} evidence (in the bluer passbands
we also trace $PA_0\approx 20\degr$ up to $R\approx 13\arcsec$,
Fig.~4). But we see the solid-body fast rotation only up to
$R\approx 5\arcsec$, and under the assumption of $i=40\degr$ the
mass contained within this radius can be estimated as $(7-10)
\cdot 10^9\,M_\sun$ -- a rather high value corresponding to the
fast circumnuclear rotation, $\omega \approx 280$~km/s/kpc.

In the center of NGC~6340 the stellar and gaseous
velocity fields differ dramatically. The lack of visible stellar
rotation must be confronted with the isophote major axis turn and
the local ellipticity maximum in the radius range
$1\arcsec - 5\arcsec$. Such combination of morphological and kinematical
properties in the center of the face-on disk galaxy may be
explained by a slight bulge triaxiality with its largest axis
aligned in $PA\approx 85\degr$ whereas the line of nodes of the galaxy
is close to $PA_0\approx 130\degr - 140\degr$. Interestingly,
the tiny dust lane noticed by \citet{cszm97} in their HST
observations of NGC~6340 goes through the nucleus
just in this direction. Obviously, the diffuse matter is going
to settle into one of the principal planes of the triaxial bulge;
in this particular case we see a quasi-polar circumnuclear gaseous
disk. The alignment of the [NII] emission isophotes (Fig.~7b) and
of the dynamical major axis of the gas velocity field (Fig.~8b)
in $PA_0\approx 115\degr - 120\degr$ proves a planar character
of the gas rotation, but not exactly in the principal plane
of the triaxial bulge. Perhaps the circumnuclear gaseous disk
is not completely stabilized yet. However the high line-of-sight
velocity gradient evidences for a rather edge-on orientation
of the rotation plane.

Circumnuclear polar gaseous disks in galaxies with triaxial bulges
or bars represent a rather new phenomenon which is not wide-known
yet. However several examples are found in nearby non-interacting
disk galaxies. We \citep{we97} have reported on the existence of a
fast-rotating gaseous polar disk with a radius of $\sim 200$ pc
in the regular Sb galaxy NGC~2841; later we have proved
that its bulge is triaxial \citep{we99}. The orthogonal planar rotations
of ionized gas and stars have also been detected by us in the
center of the lenticular galaxy NGC~7280 \citep{we7280};
this galaxy also possesses an intermediate-scale bar. Another
example of the circumnuclear polar gaseous disk is known in
NGC~253 \citep{ag96}. Qualitative arguments \citep{sw94}
evidence that an existence of circumnuclear polar gaseous disks
well inside triaxial stellar structures may be a result of
intrinsic dynamical evolution; an accretion of external gas is not
quite necessary in such a case. However, an extended stellar-and-gaseous
disk with a randomly inclined rotation axis in the axisymmetric
galaxy, such as NGC~524, seem to require external
gas (and stars?) accretion.

Another important question which we try to answer for some time is if
a chemically distinct nucleus (core) and a circumnuclear stellar disk
are the same thing or are they different substructures though perhaps
evolutionarily related? In elliptical galaxies the former variant
is discussed for several years \citep{sb95,car97}. But in some spiral
galaxies we clearly resolve the circumnuclear stellar disks and
simultaneously cannot resolve their chemically distinct nuclei.
Perhaps, the brightest example is M~31: the size of its chemically
distinct entity is restricted by $R \le 4\arcsec$ whereas the
circumnuclear stellar disk extends up to $R \approx 30\arcsec$
\citep{silbv98}. Though less convincing due to the worse spatial
resolution, a similar situation appears to be present in NGC~524.
The kinematically traced inclined stellar disk is seen at
$R=4\arcsec$, but the effect of the chemically distinct nucleus
is already negligible at this radius. So we would prefer to separate
the chemically distinct nucleus and the circumnuclear stellar disk
in this galaxy. Rather we would associate the chemically distinct
nucleus in NGC~524 with its "core" that has been found from WFPC HST
observations of the central luminosity profile \citep{hst2}. The
radius of the photometrically distinct core in NGC~524 is
$0\farcs 3$ that cannot be resolved in our observations; but as
it is detectable from photometry, it may affect the spectral
characteristics of the galactic nucleus if the stellar population
of the core is different from the stellar population of the bulge.
As for NGC~6340, we do not know if this galaxy has a separate
circumnuclear stellar disk -- not identifying by its kinematics,
it may be only face-on. But a photometric core detected from
the HST data is also present in NGC~6340; its radius is
$0\farcs 5$ \citep{cs3}, so we would see it as an unresolved
issue.

\acknowledgements
I thank the astronomers of the Special Astrophysical Observatory
Drs. V. L. Afanasiev, S. N. Dodonov, V. V. Vlasyuk,
and Mr. Drabek for supporting the observations at the 6m telescope.
I am also grateful to the post-graduate student of the Special
Astrophysical Observatory
A. V. Moiseev for the help in preparing the manuscript.
The 6m telescope is operated under the financial support of
Science Ministry of Russia (registration number 01-43).
During the data analysis I have
used the Lyon-Meudon Extragalactic Database (LEDA) supplied by the
LEDA team at the CRAL-Observatoire de Lyon (France) and the NASA/IPAC
Extragalactic Database (NED) which is operated by the Jet Propulsion
Laboratory, California Institute of Technology, under contract with
the National Aeronautics and Space Administration.
This research has made use of the La Palma Archive. The telescope
JKT is operated on the island of La Palma by the Royal
Greenwich Observatory in the Spanish Observatorio del Roque de los
Muchachos of the Instituto de Astrofisica de Canarias.
The work is partly based
on observations made with the NASA/ESA Hubble Space Telescope, obtained
from the data archive at the Space Telescope Science Institute, which is
operated by the Association of Universities for Research in Astronomy,
Inc., under NASA contract NAS 5-26555. The work
was supported by the grant of the Russian Foundation for Basic
Researches 98-02-16196, by the grant of the President of Russian
Federation for young Russian doctors of sciences 98-15-96029
and by the Russian State Scientific-Technical
Program "Astronomy. Basic Space Researches" (the section "Astronomy").

\clearpage

\figcaption[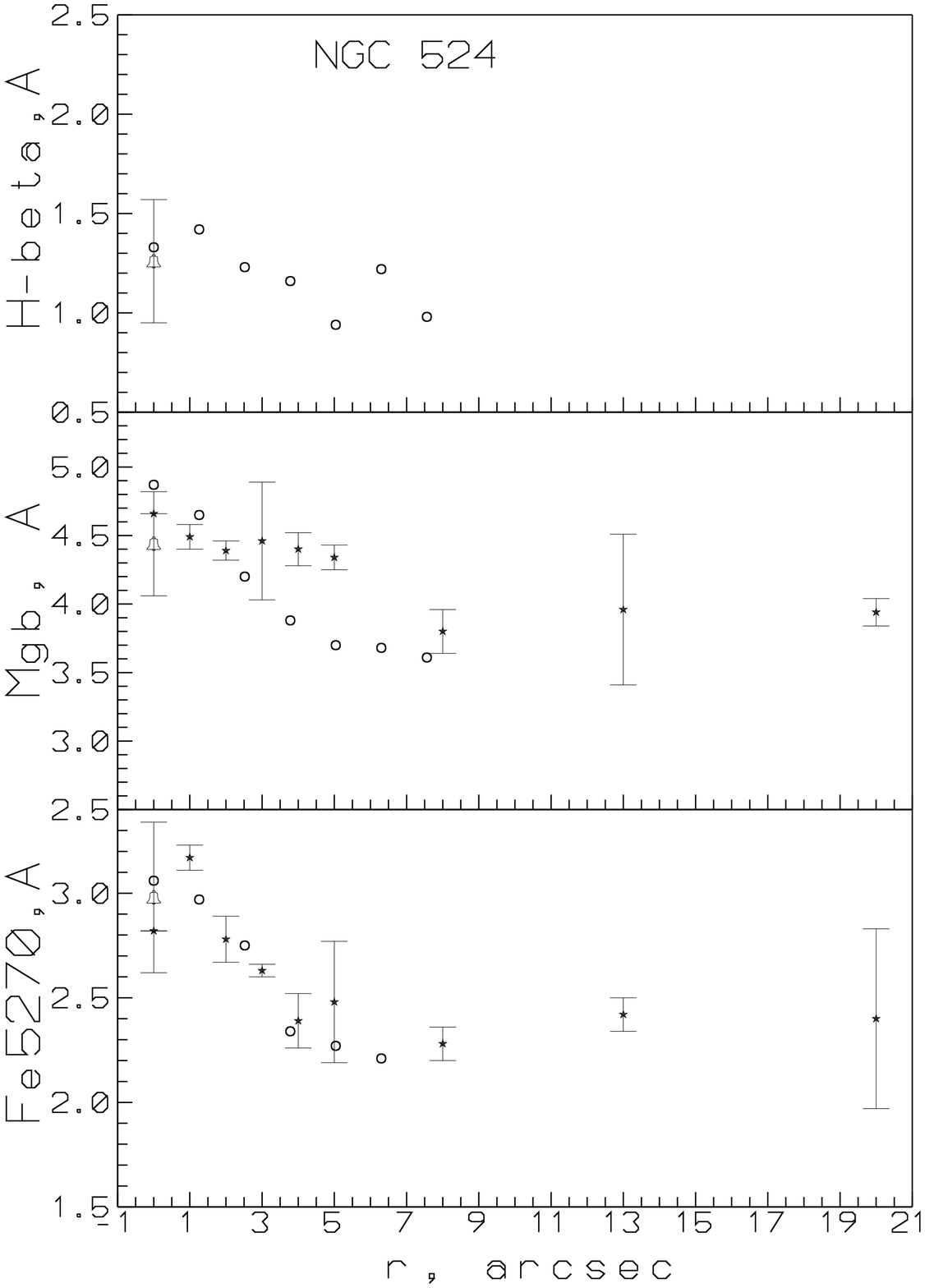]{NGC 524: Radial profiles of the absorption-line
indices $\mbox{H}\beta$, Mgb, and Fe5270; open circles
present our MPFS data, the bell with the error bar is the data
of Trager et al. (1998). The asterisks show the long-slit data
from the ISIS WHT (La Palma Archive) not corrected for the stellar
velocity dispersion and not reduced to the standard Lick system.
\label{fig1}}

\figcaption[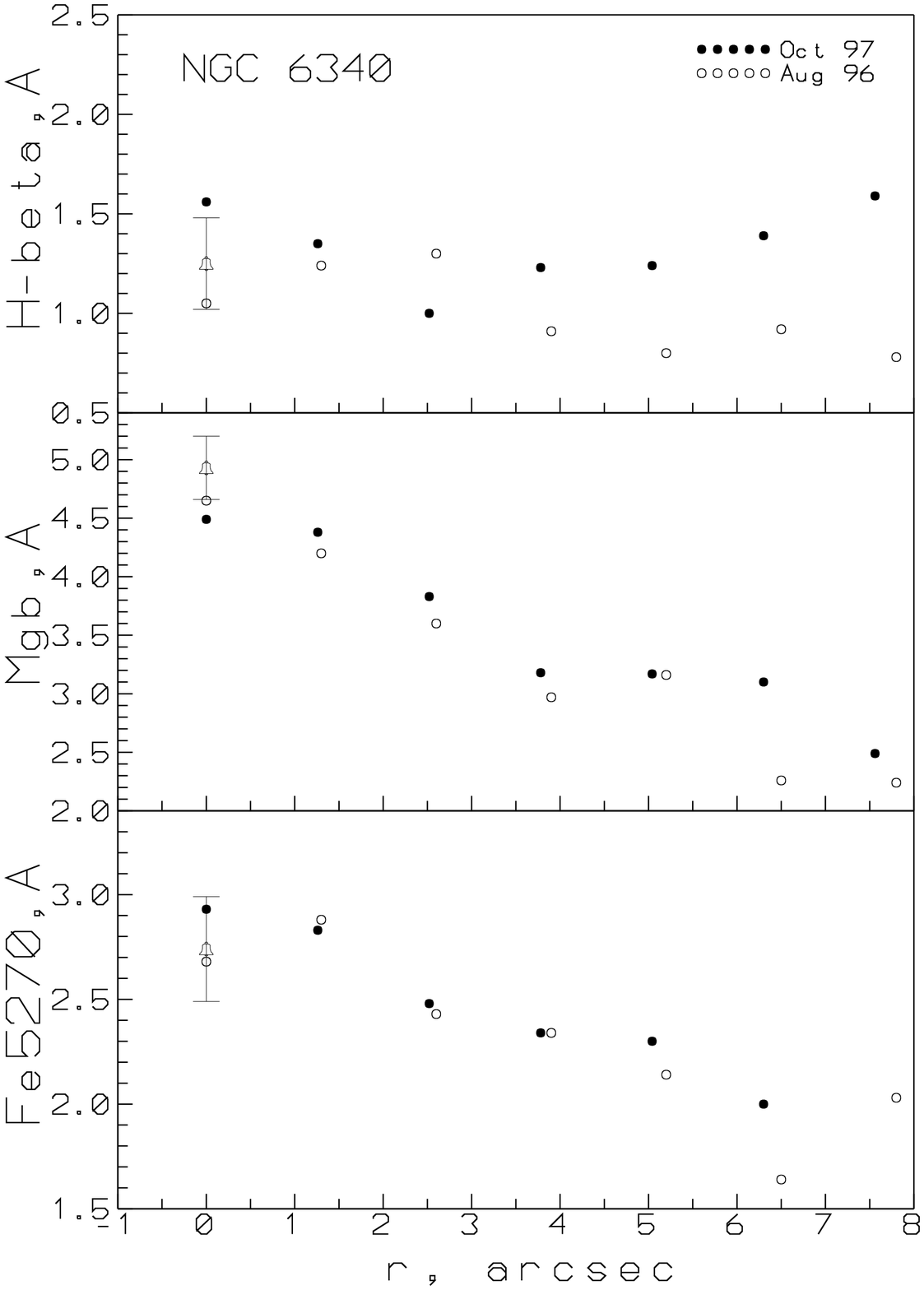]{NGC 6340: Radial profiles of the absorption-line
indices $\mbox{H}\beta$, Mgb, and Fe5270; filled and open circles
present our MPFS data, the bell with the error bar is the data
of Trager et al. (1998). \label{fig2}}

\figcaption[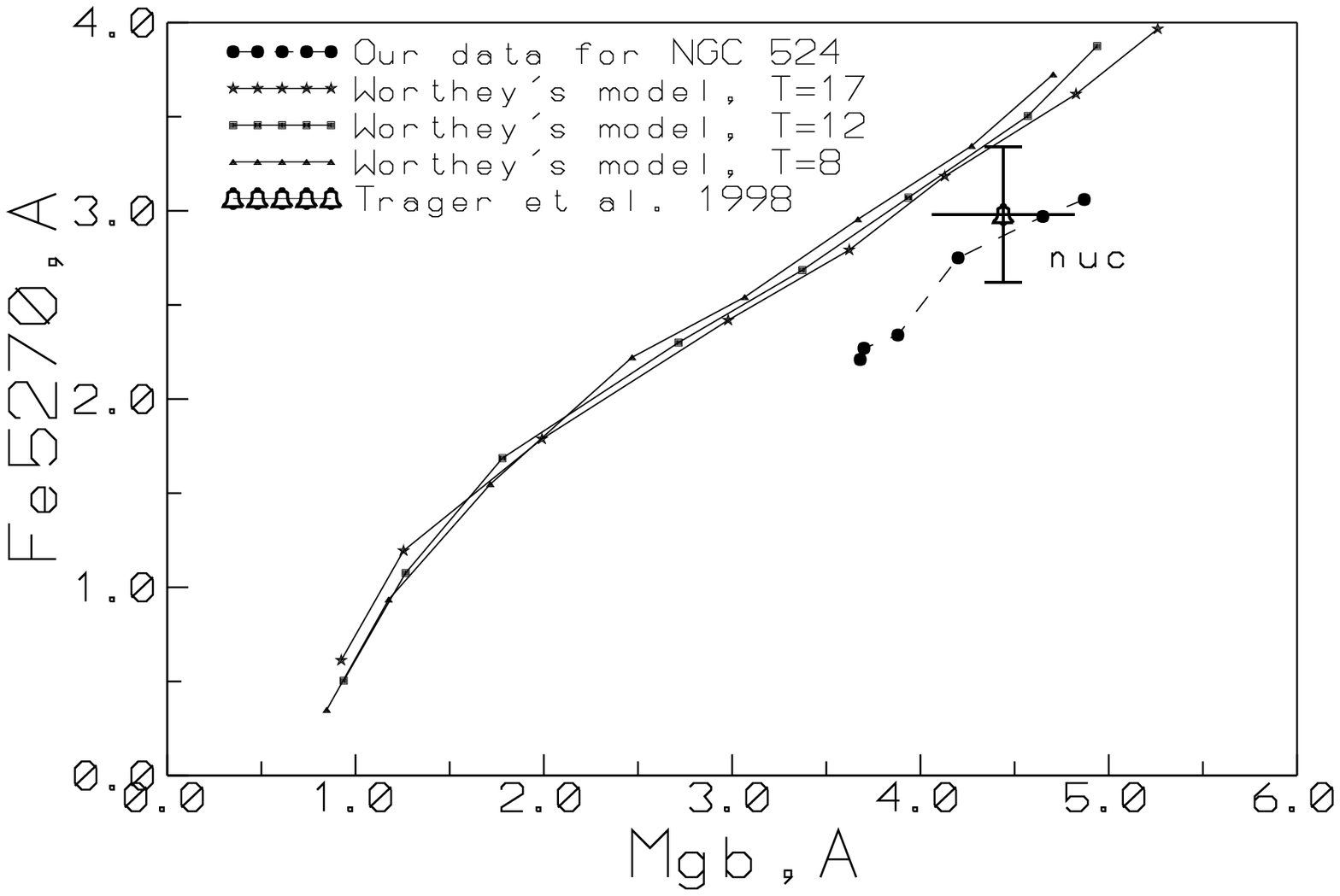,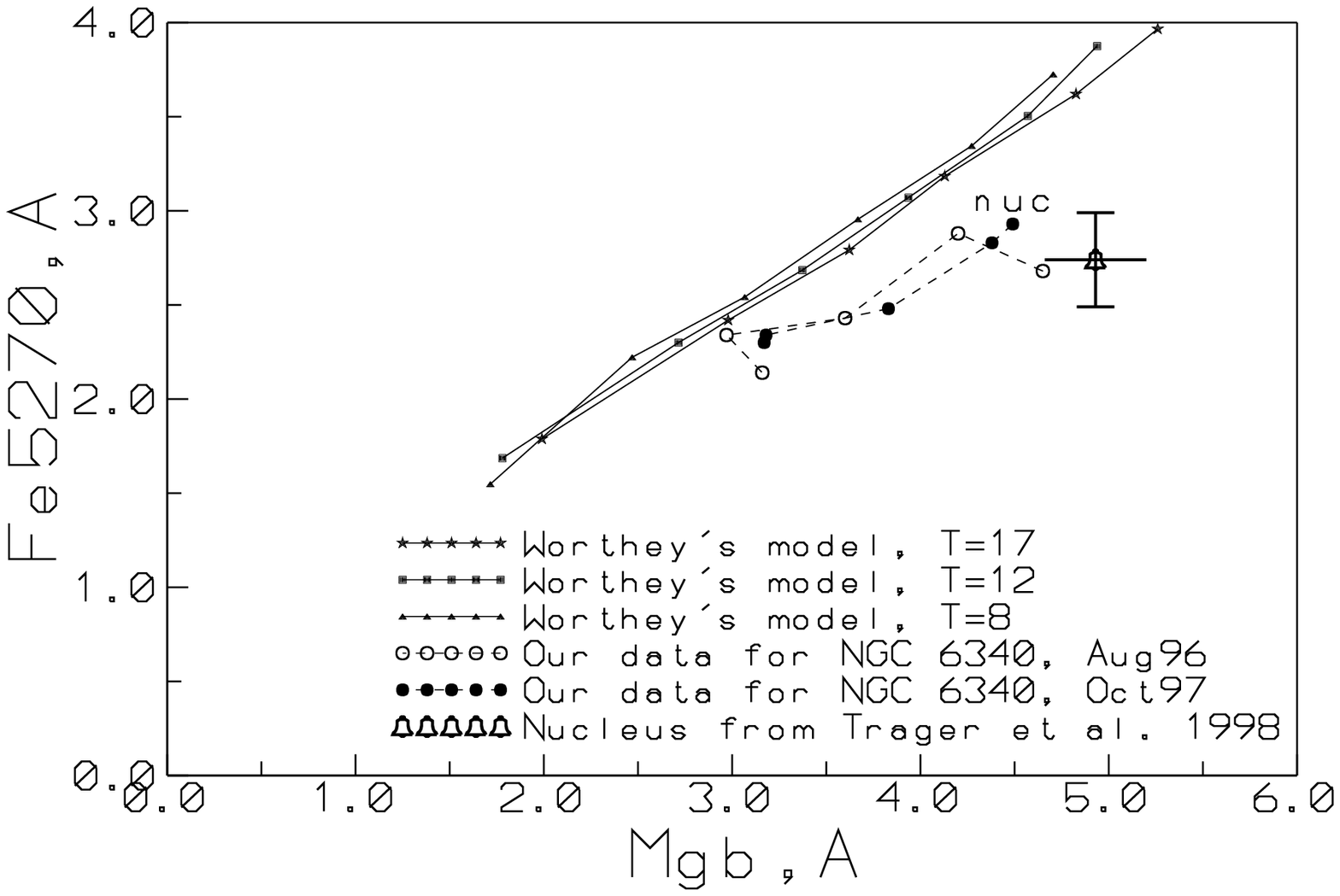]{The diagrams (Fe5270, Mgb): {\it a} --
for NGC~524, {\it b} -- for NGC~6340.
The measurements are azimuthally averaged and taken along the radius
with the step of $1\farcs 3$.
The nuclear measurements from Trager et al. (1998) are plotted for
comparison with their error bars.
The ages of the Worthey's (1994) models for [Mg/Fe]=0
are given in billion years. \label{fig3}}

\figcaption[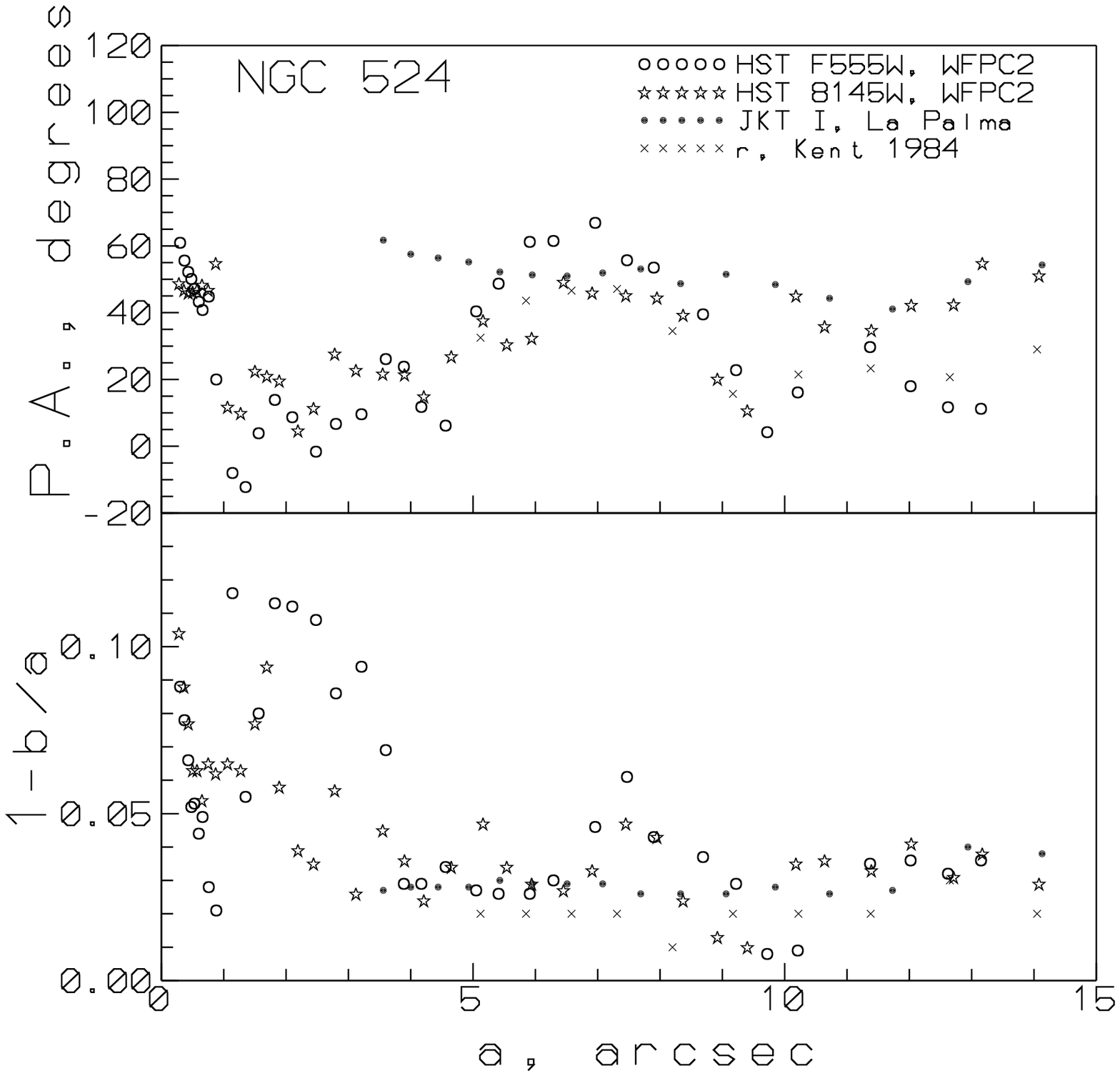]{Radial variations of the isophote morphological
characteristics in the center of NGC~524 according to the HST data
and to the La Palma data. \label{fig4}}

\figcaption[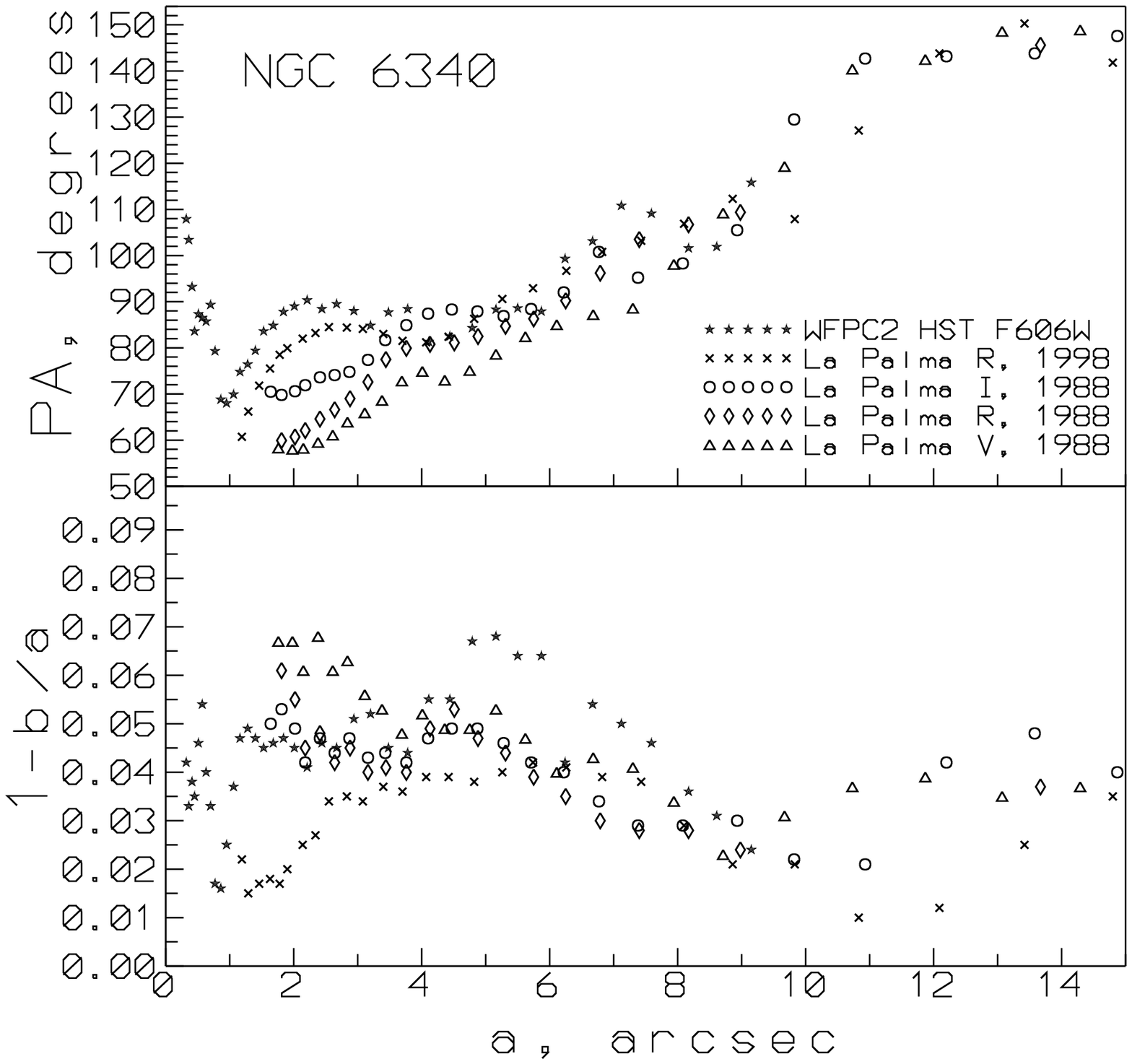]{Radial variations of the isophote morphological
characteristics in the center of NGC~6340 according to the HST data
and to the La Palma data. \label{fig5}}

\figcaption[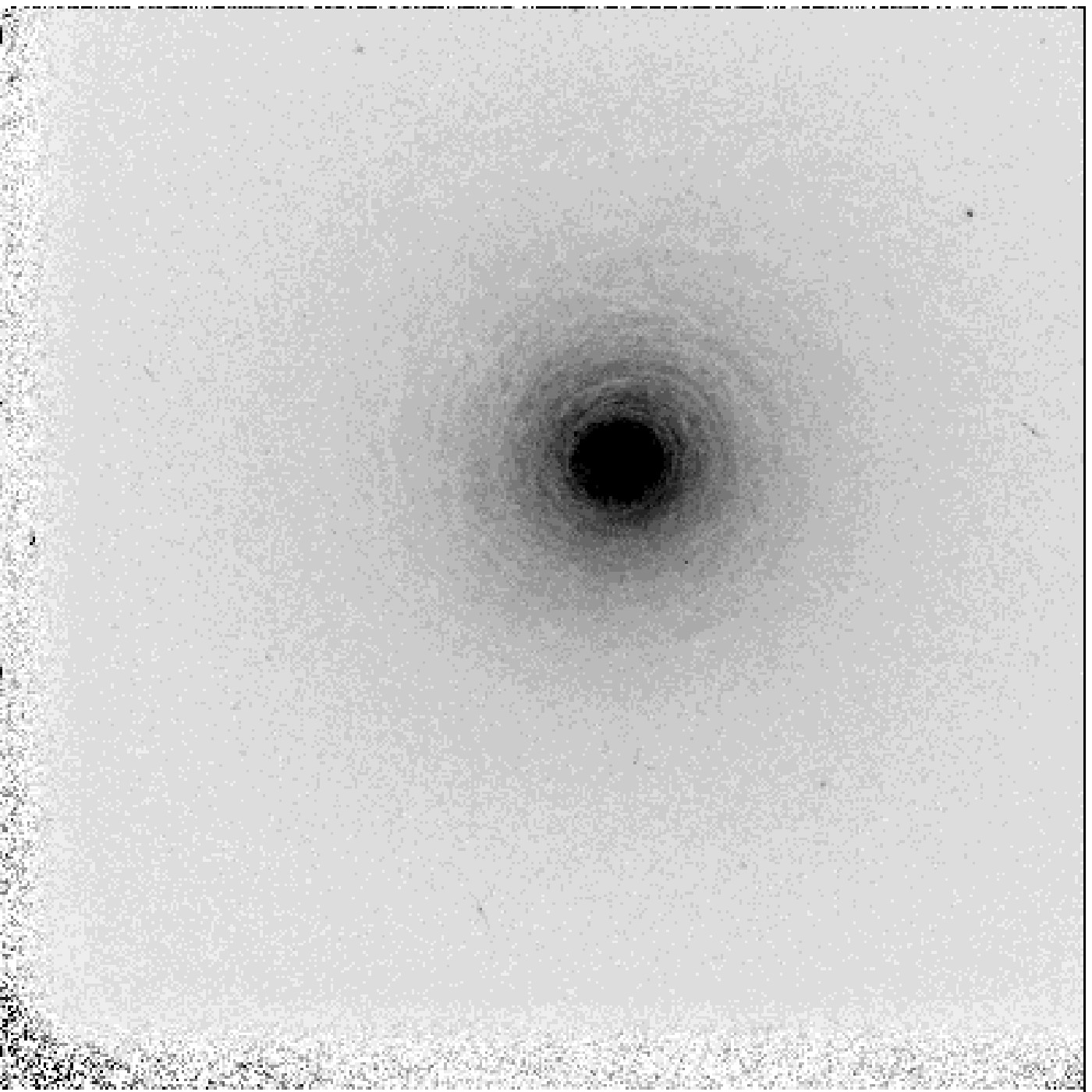,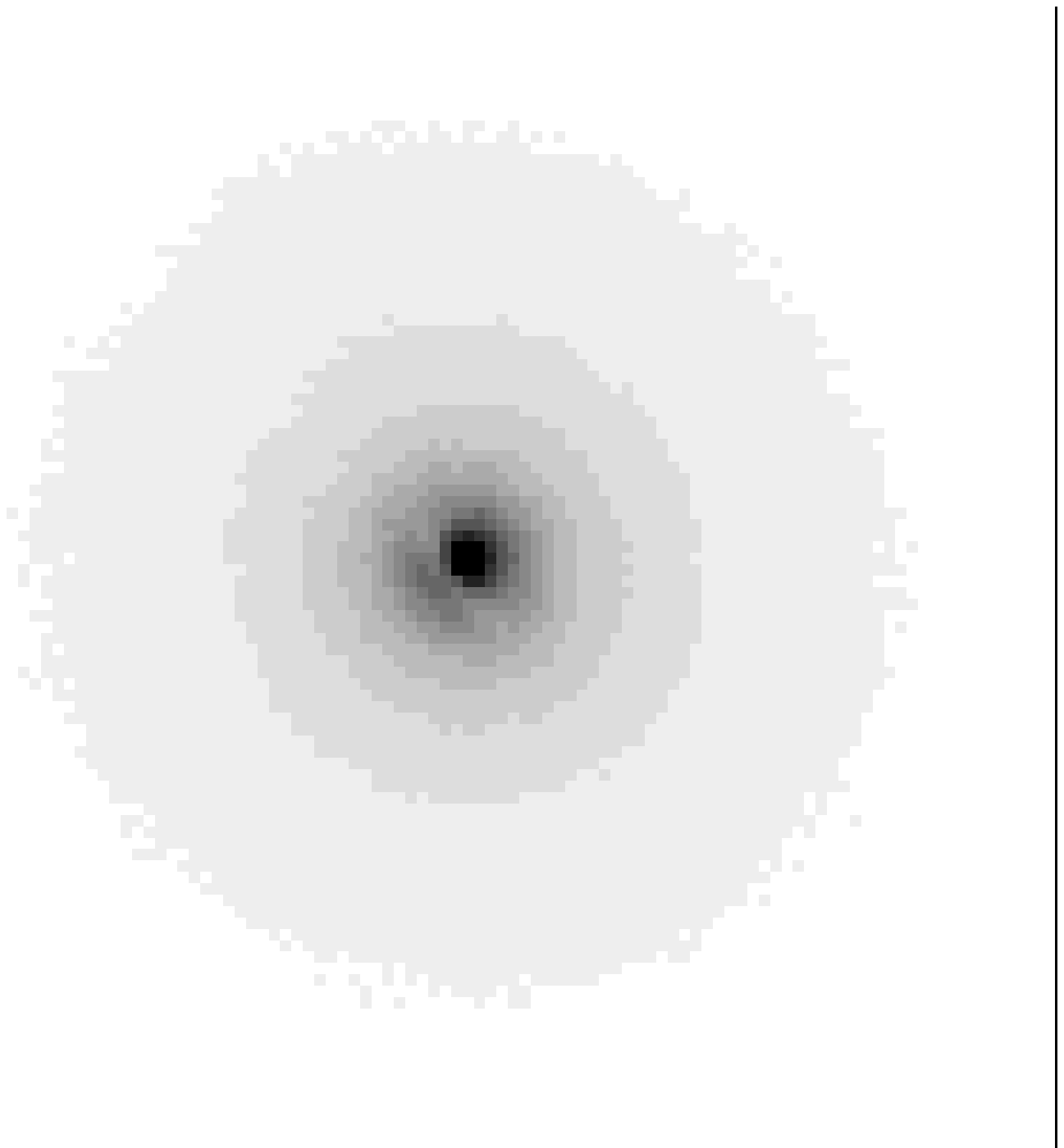]{The direct views of the images of
NGC~524 ({\it a}) and
NGC~6340 ({\it b}) obtained by the HST WFPC2 in the visual passbands.
The sizes of the maps shown are $36\arcsec \times 36\arcsec$ and
$4\farcs 5 \times 4\farcs 5$, respectively,
the orientations are $PA(top)=-68\degr$ and $PA(top)=-129\degr$.
\label{fig6}}

\figcaption[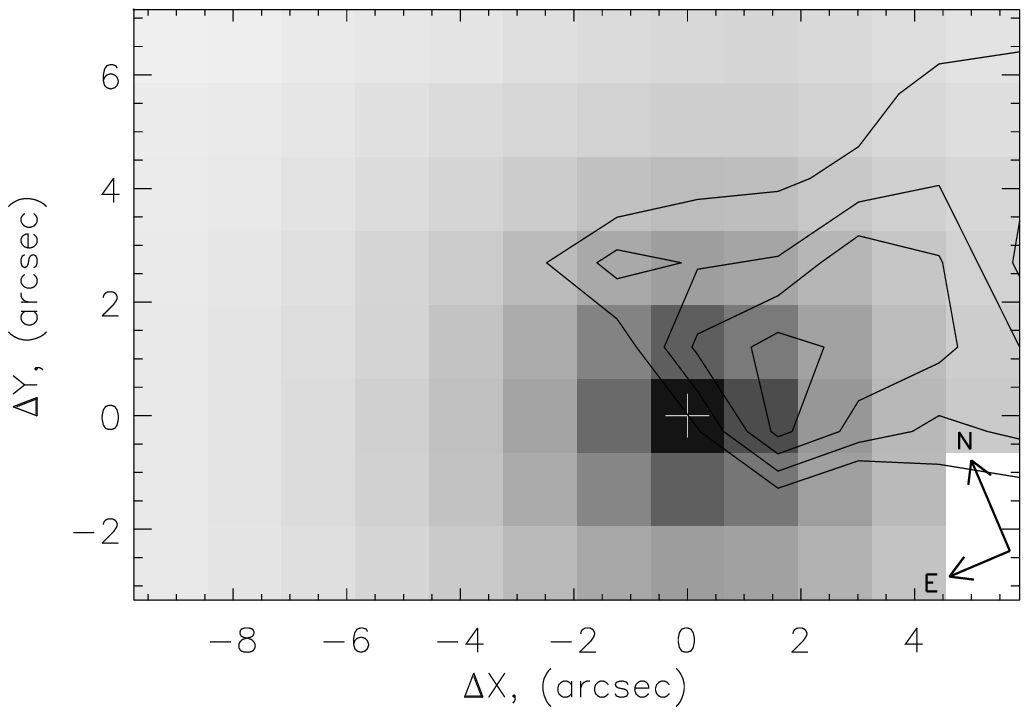,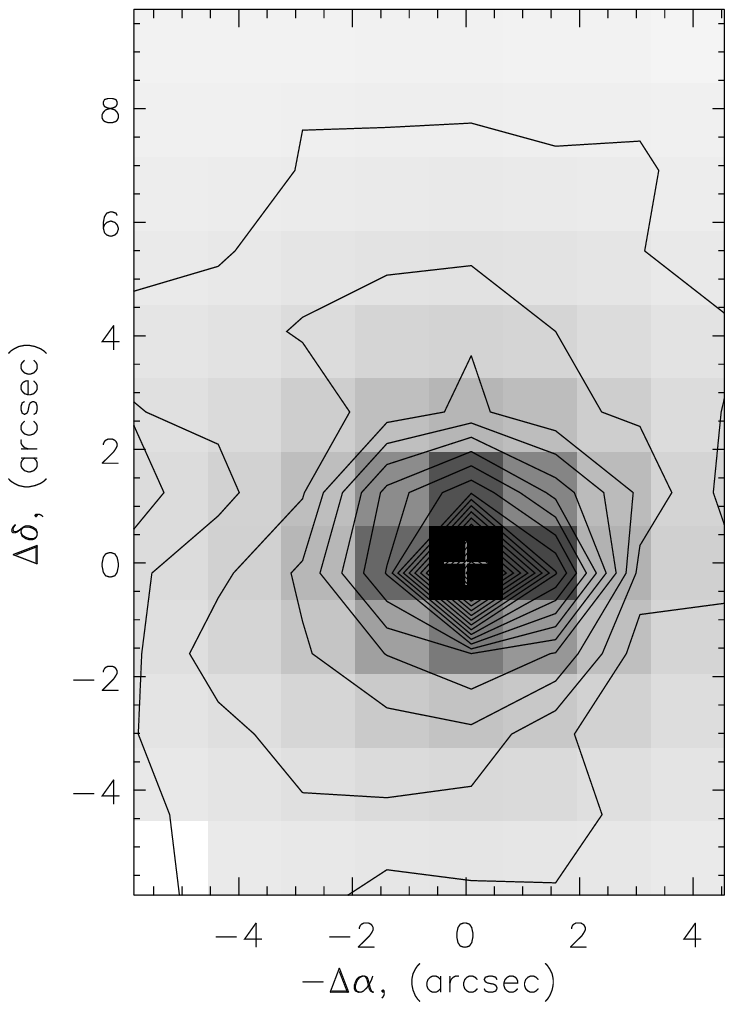]{The surface brightness distributions
for the emission
line [NII]$\lambda$6583 (isolines) superimposed on the gray-scaled
red continuum map for NGC~524 ({\it a}) and NGC~6340 ({\it b}).
The maps are direct, the orientations are $PA(top)=-23\degr$ for
NGC~524 and $PA(top)=0\degr$ for NGC~6340.
The brightness units are arbitrary. \label{fig7}}

\clearpage

\figcaption[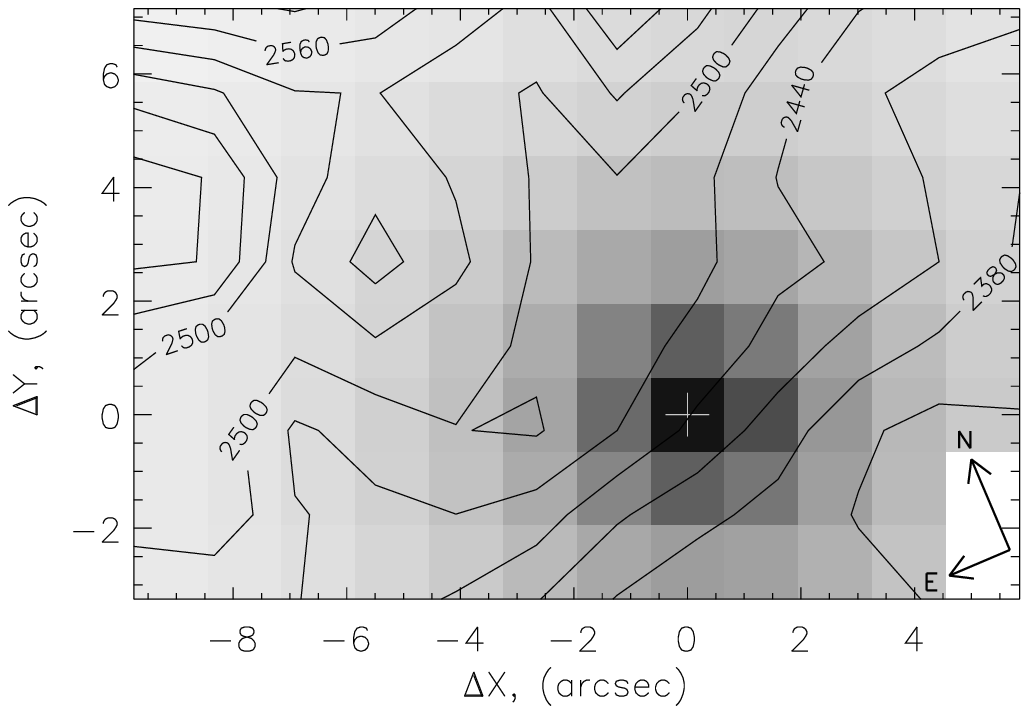,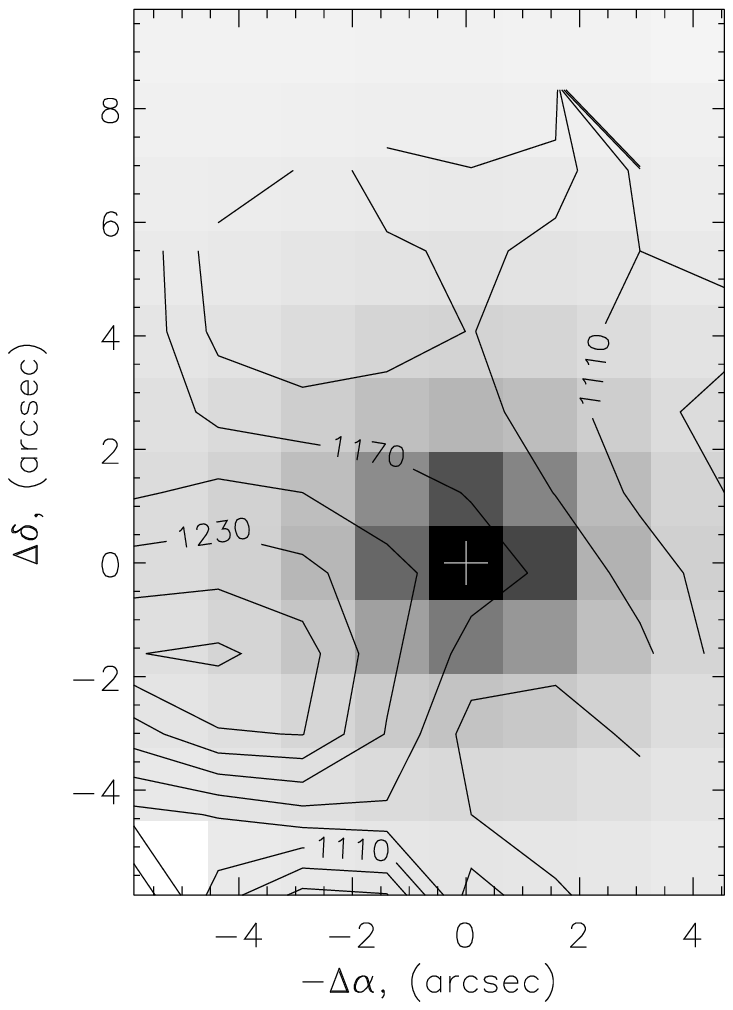]{Isovelocities for the ionized gas in
the centers of NGC~524 ({\it a}) and NGC~6340({\it b}).
The maps are direct, the orientations are $PA(top)=-23\degr$ for
NGC~524 and $PA(top)=0\degr$ for NGC~6340. The positions of photometric
centers are shown by the white crosses. \label{fig8}}

\figcaption[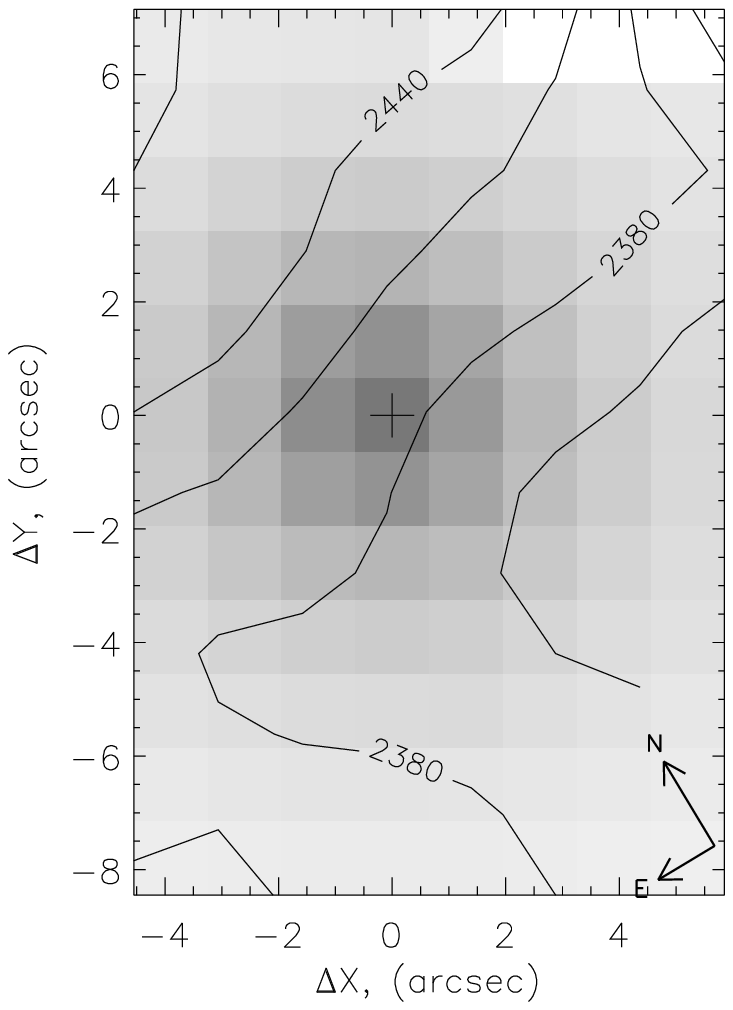]{Stellar isovelocities in the centers of NGC~524.
The map is direct, $PA(top)=-31\degr$, the position of the photometric
center is shown by the cross. \label{fig9}}

\figcaption[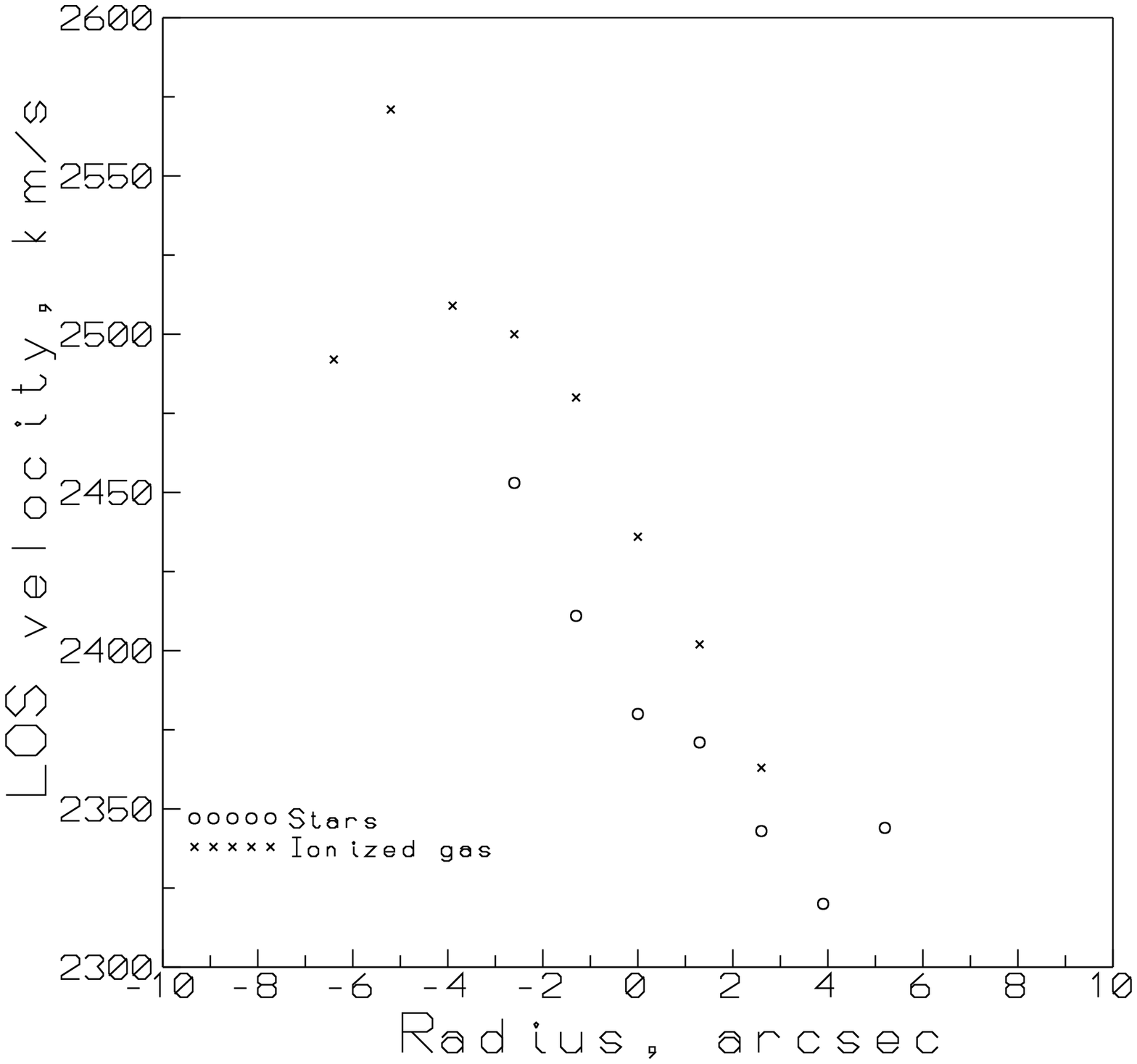,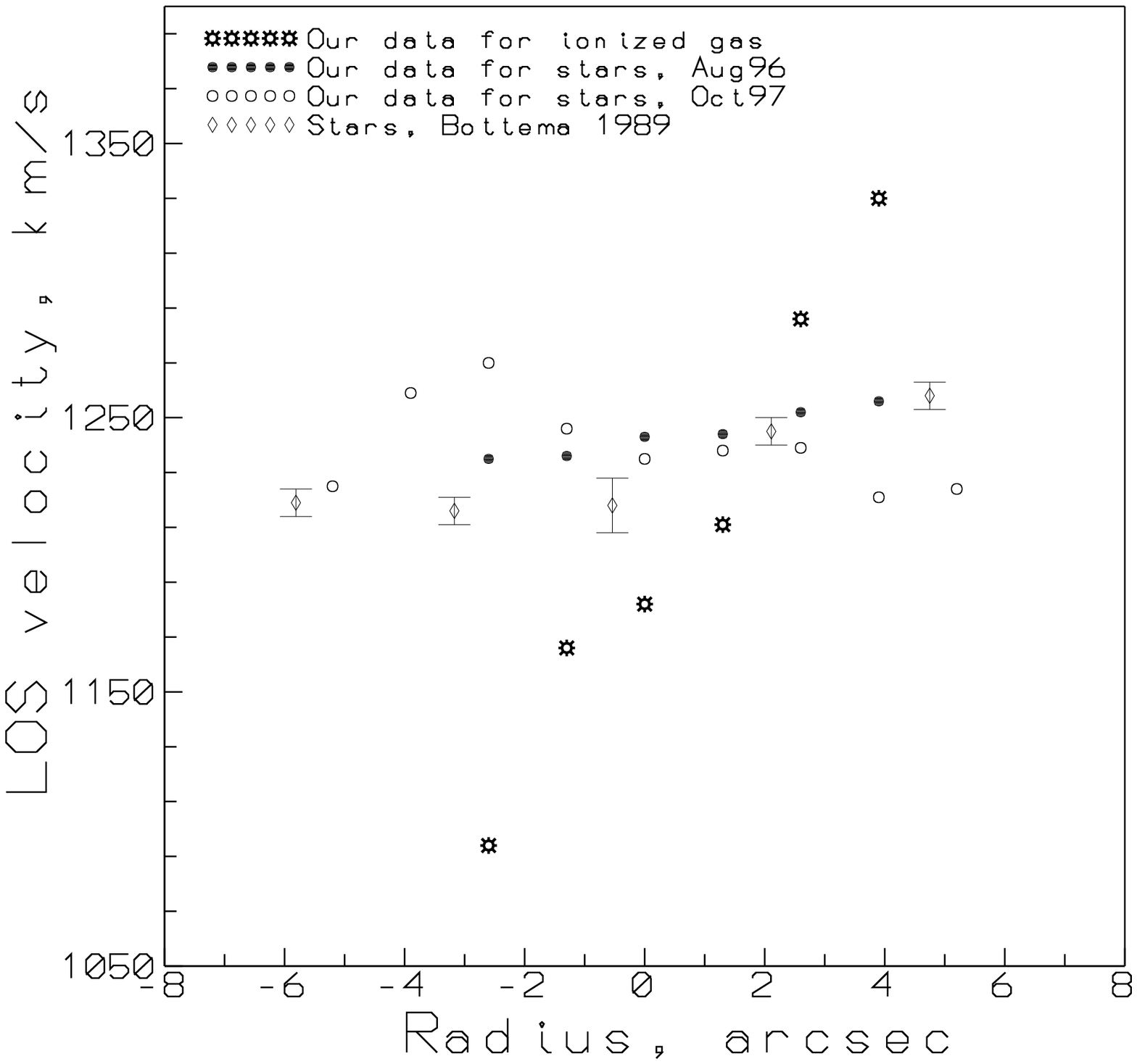]{Major-axis line-of-sight velocity
profiles for the
stellar and gaseous components in the centers of NGC~524 ({\it a}),
$PA(cut)=20\degr$, and NGC~6340 ({\it b}), $PA(cut)=130\degr$.
\label{fig10}}

\end{document}